\documentclass[twocolumn]{emulateapj}
\usepackage{apjfonts}
\usepackage{soul}

\slugcomment{}

\begin{document}


\title{Deep multi-telescope photometry of NGC 5466. II. The radial
 behaviour of the mass function slope\altaffilmark{1}}

\author{G. Beccari\altaffilmark{2},
E. Dalessandro\altaffilmark{3},
B. Lanzoni\altaffilmark{3},
F. R. Ferraro\altaffilmark{3},
M. Bellazzini\altaffilmark{4},
A. Sollima\altaffilmark{4}}

\altaffiltext{1}{Based on data acquired using the Large Binocular Telescope (LBT).  The LBT is an international collaboration among institutions in the United States, Italy and Germany. LBT Corporation partners are: The University of Arizona on behalf of the Arizona university system; Istituto Nazionale di Astrofisica, Italy; LBT Beteiligungsgesellschaft, Germany, representing the Max-Planck Society, the Astrophysical Institute Potsdam, and Heidelberg University; The Ohio State University, and The Research Corporation, on behalf of The University of Notre Dame, University of Minnesota and University of Virginia.}
\affil{\altaffilmark{2} European Southern Observatory, Karl--Schwarzschild-Strasse 2,
85748  Garching bei M\"unchen, Germany, gbeccari@eso.org}
\affil{\altaffilmark{3} Dipartimento di Fisica e Astronomia, Universit\`a 
degli Studi di Bologna, viale Berti Pichat 6/2, I--40127 Bologna, Italy}
\affil{\altaffilmark{4} INAF--Osservatorio Astronomico 
di Bologna, via Ranzani 1, I--40127 Bologna, Italy}

\date{27 August, 2015}

\begin{abstract}
 
We use a combination of data acquired with the Advanced Camera for
Survey (ACS) on board the Hubble Space Telescope and the Large
Binocular Camera (LBC-blue) mounted on the Large Binocular Telescope,
to sample the main sequence stars of the globular cluster NGC~5466 in the mass range
$0.3<M/M_\odot<0.8$.  We derive the cluster's Luminosity Function 
in several radial regions, from the center of the cluster out to the tidal radius. 
After corrections for incompleteness and field-contamination, this has been 
compared to theoretical Luminosity Functions, obtained by multiplying
a simple power law Mass Function in the form dN/dm$ \propto m^{\alpha}$
by the derivative of the mass-luminosity relationship of the best-fit isochrone.
We find that $\alpha$ varies from -0.6 in the core region to -1.9 in the outer region. 
This fact  allows us to observationally prove that the stars in NGC 5466 have experienced 
the effects of mass segregation.
We compare the radial variation of $\alpha$ from the center out to 5 core radii (r$_c$)
in NGC 5466  and the globular cluster M10, finding that the gradient of $\alpha$ in the first 5r$_c$ is more than a
factor of 2 shallower in NGC 5466 than in M10, in line with the differences in the
clusters' relaxation timescales. NGC 5466 is dynamically younger than M10, with two-body
relaxation processes only recently starting to shape the distribution
of main sequence stars.  This result fully agrees
with the conclusion obtained in our previous works on the radial
distribution of Blue Straggler Stars, further confirming that this can
be used as an efficient clock to measure the dynamical age of stellar
systems.
\end{abstract}

\keywords{stars: luminosity function, mass function--- binaries: general --- globular clusters:
  general --- globular clusters: individual(NGC 5466)}

\section{Introduction} 
\label{intro}
Globular Clusters (GCs) are among the most fascinating and intensively
studied objects in the Galaxy. They are typically populated by millions of stars,
whose age, distance and chemical abundance are well known, thus making
GCs the ideal benchmark to study stellar and dynamical evolution and
to understand how these two apparently independent evolutionary
channels can influence each other.

While the main engine of stellar evolution are stellar thermonuclear
reactions, the long-term dynamical evolution of GCs is driven by
two-body relaxation.  The typical time-scale in which two-body
processes take place in stellar systems depends on their masses and radii
\citep[][]{spi87} and for GCs it is typically significantly shorter
\citep[1-2 Gyr;][]{me97} than their age. Therefore GCs may have
experienced basically all the phases of dynamical evolution.  Indeed
they survive the early and violent expansion triggered by primordial
gas expulsion and mass loss due to stellar evolution, then they evolve
toward higher central densities, eventually reaching core collapse
while loosing stars through the boundary set by the tidal field of
their host galaxy \citep[][]{hh03}.

Because of two-body relaxation, heavier objects tend to sink toward
the cluster centers (mass segregation), while less massive stars are
forced toward more external orbits.  Hence, one possible observational
approach to trace the dynamical state of star clusters is to look at
the radial variations of the luminosity function (LF) and$/$or mass
function (MF) of main sequence (MS) stars~\citep[see e.g.][]{da82,ma01,al02,ko04,de07}.  This method only allows to
look for the effect of mass segregation in a range of masses between
$\sim0.8M_\odot$ (the present-day mass of stars located at the MS
Turn-Off point of GCs) down to the mass corresponding to the limiting 
magnitude of observational data. Because of the
effect of mass segregation, the slope $\alpha$ of the MF is expected
to become steeper as the distance from the cluster center
increases~\citep[e.g.][]{ro99,lee03,lee04,an04}. For example in \citet[][]{be10} we studied the radial
distribution of the MF slope in M10. We found $\alpha$ to drop from
0.7 in the center to -0.9 in the external regions. Supported by N-body
simulations, we interpreted this radial change of the MF slope as a
clear sign of mass segregation in M10.  The obvious weakness in this
approach is the need of very deep and accurate photometric
measurement, even in the very central regions of the clusters where
stellar densities seriously challenge the spatial resolution
capabilities of even space-based observations.

Similarly, internal dynamics can be probed by means of massive ``test
particles'' like Blue Straggler Stars (BSSs), binaries and Millisecond
Pulsars \citep[][]{gu98,fe01,fe03,di05,he06}.  Among them BSSs have been widely used
for this purpose since they are relatively bright (and hence easily observable) 
and they are a typical population of any GC.  \citet[][]{fe12} have shown that
the BSS radial distribution can be efficiently used to rank clusters
according to their dynamical age and defined an empirical tool (``the
dynamical clock'') able to extract information about the dynamical
state of GCs on the basis of only the position of the observed minimum
in the BSS radial distribution.  Three families have been identified in
this way. {\it Family I} clusters are dynamically young, not showing
significant signs of mass segregation and having a flat BSS radial
distribution (when properly compared to normal stars or to the sampled
light).  On the contrary, {\it Family II} and {\it Family III} GCs are
classified as dynamically intermediate and old, as they show bimodal
and monotonically decreasing BSS radial distributions,
respectively. As expected, the large majority of the GCs studied so
far are in a dynamically evolved state, with only a few notable
exceptions \citep[$\omega$ Centauri, NGC 2419, Palomar 14, NGC 6101,
  Arp 2 and Terzan 8;][]{fe06, da08, da15, be11, sa12}.  In general a
good agreement has been found among different indicators (see for
example Beccari et al.  2006 and 2013 for the cases of M62 and NGC
5466, respectively; Dalessandro et al. 2008 and Bellazzini et al. 2012
for NGC 2419). In particular, we have recently performed a detailed
analysis of the dynamical state of NGC 6101 \citep{da15}, studying
three different dynamical indicators (the BSS and the binary radial
distributions and the radial variation of the LF and MF) and we consistently found
a significant lack of mass segregation in this system.  Opposite
results from different indicators are obtained, instead, for the case
of Palomar 14.  In fact, based on the analysis of the radial
distribution of BSSs \citet[][]{be11} concluded that this cluster is
not relaxed yet, while \citet[][]{fr14} revealed a non-negligible
variation of the MF slope across the cluster
inner regions. This apparent discrepancy between the two results
can be reconciled with the hypothesis that the cluster was either 
primordially mass segregated and/or used to be significantly more compact in the past~\citep[][]{fr14}.
Palomar 14 is one of the most remote GC in the Galaxy ($d\sim66$ kpc), which makes it very 
hard to sample the MS at masses below 0.5 M$_{\odot}$ even with 8m class telescopes. 
This fact once more suggests that the BSSs are a privileged sample of test particles to study the cluster's dynamics
being these massive stars more than 6 magnitudes brighter 
with respect to the low mass end of the MS.

In \citet[][hereafter Paper I]{be13} we studied the BSS population of
the Galactic GC NGC 5466, finding that it shows a bimodal radial
distribution with a mild central peak and a minimum located at only $\sim2.5r_c$,
where $r_c=72''$ is the cluster's core radius~\citep[][]{mio13}. In
the framework of the ``dynamical clock'' (Ferraro et al. 2012), we
interpreted this feature in terms of a relatively young dynamical
age. Interestingly, we found that also the radial distribution of
binary stars seems to display a bimodal behaviour, with the position
of the minimum consistent with that of BSSs. In this paper we present
the first analysis of the LF and MF along the entire radial extension
of NGC 5466, obtained by combining deep high-resolution Hubble Space
Telescope (HST) observations and ground-based Large Binocular
Telescope (LBT) data. We compare these new results with those obtained
in Paper I and we discuss their implication for our understanding of
the dynamical state of this system.  The paper is structured as
follows: in Section \ref{sec_obs} we describe the observations and
data-analysis procedure, in Section \ref{sec_lum} we derive the LF and
the slope of the MF of NGC 5466 at different distances from the
cluster center and in Section \ref{sec_disc} we discuss the obtained
results in the framework of the dynamical evolution of the cluster.

\section{Catalogs and photometric completeness}
\label{sec_obs}
This work is based on the combination of deep high-resolution
observations obtained with the HST Advanced Camera for Survey (ACS),
and wide-field images acquired with the Large Binocular Camera (LBC)
mounted at the LBT.  The adopted reduction procedures are described in
Paper I, where the photometric data-set used here is referred as the
``Deep Sample'' (see also Table 1). 

\begin{figure} \centering
\plotone{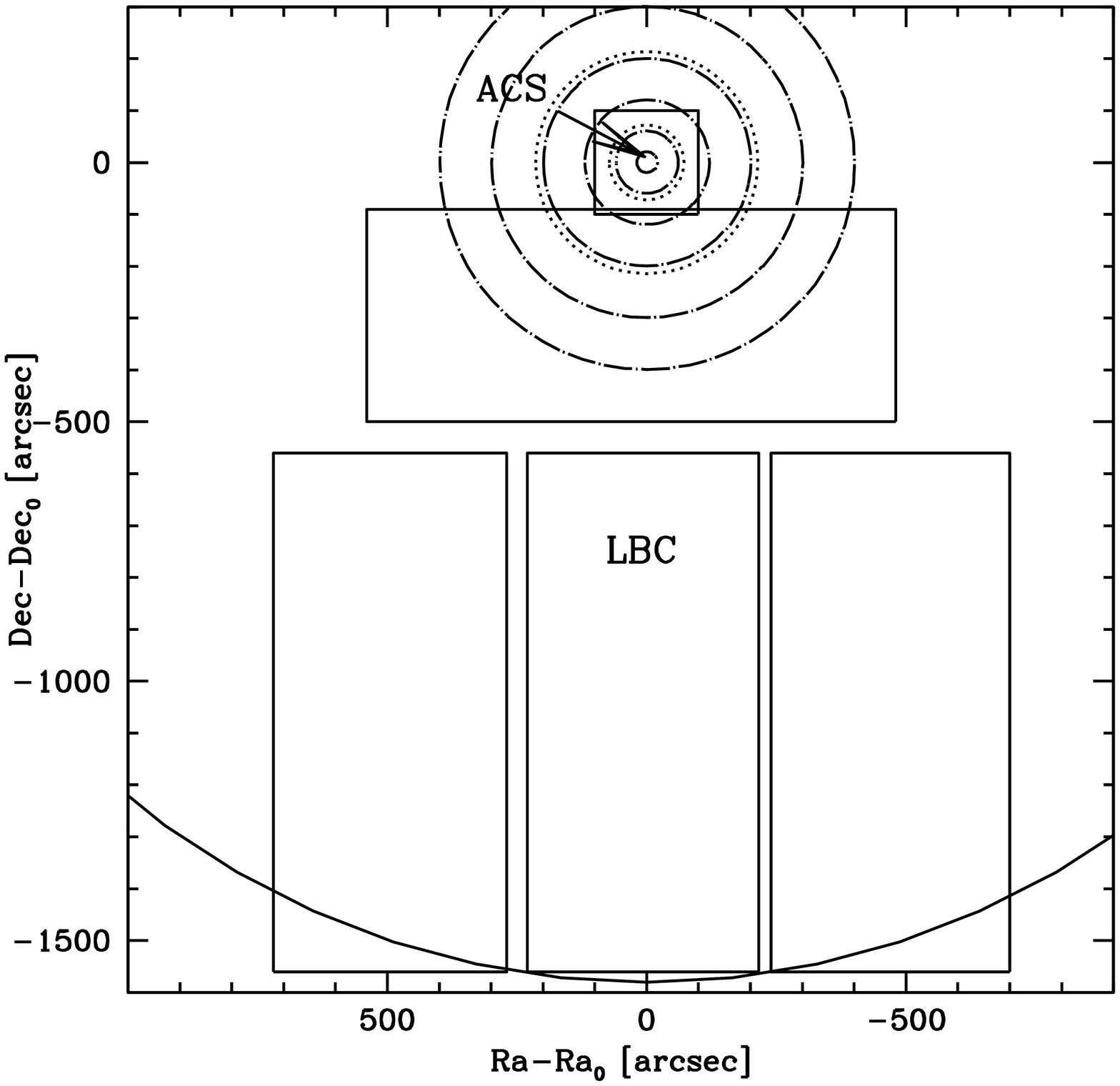}
\caption{Map of the combined FOV of the data-sets used to study the LF
  of NGC 5466 (this also corresponds to the ``Deep Sample'' used to
  study the cluster BSS and binary fractions in Paper I).  The solid
  square corresponds to the ACS FOV, while the four rectangles mark
  the FOV of the LBC sample.  The two dotted circles indicate the
  position of the core and half-mass radii ($r_c=72\arcsec$ and
  $r_h=214\arcsec$, respectively), while the large solid circle
  indicates marks the location of the cluster's tidal radius
  ($r_t=1580\arcsec$).  The cluster center is taken from Paper I 
  and physical parameters are derived from \citet[][]{mio13}.  
  The dashed-dotted circles mark the
  radial regions considered in the present study for the analysis of
  the cluster LF, with the last annulus extending out to the tidal
  radius.}
\label{maps}
\end{figure}

Briefly, we used the ACS images to sample the cluster MS in the
$F606W$ and $F814W$ bands from the Turn-Off (V$\sim$20.5) down to
V$\sim27$, in the first $\sim120\arcsec$ from the cluster center. The
deep LBC data allowed us to obtain a photometric catalog in the $B$
and $V$ bands sampling the MS down to comparable magnitudes in an area
extending out to the cluster's tidal radius ($r_t=1580\arcsec$;
Miocchi et al. 2013).  In Figure \ref{maps} we show the position of
the field of view (FOV) of the HST and the LBT data-sets with respect
to the cluster center derived in Paper I.

The data reduction of the whole data-set was performed through a
standard point spread function (PSF) fitting procedure, by using
DAOPHOTII/ALLSTAR \citep{st87,st94}.  We independently calibrated the
ACS instrumental magnitudes into the VEGAMAG system adopting the
standard procedure described in \citet[][]{sir05}\footnote{We used the
  newly zero points values which are available at the STScI web pages:
  http://www.stsci.edu/hst/acs/analysis/zeropoints}.  We transformed
the instrumental $B$ and $V$ magnitudes of the LBC sample into the
Johnson/Kron-Cousins standard system by means of more than 200 stars
in common with a photometric catalog previously published by
\citet{fek07}.  Finally the $F606W$ filter was transformed into the
Johnson $V$ magnitude using the stars in common between the ACS and
the LBC catalogs.  This provides us with a homogenous $V$ magnitude
scale in common between the two data-sets.

\begin{figure}
\plotone{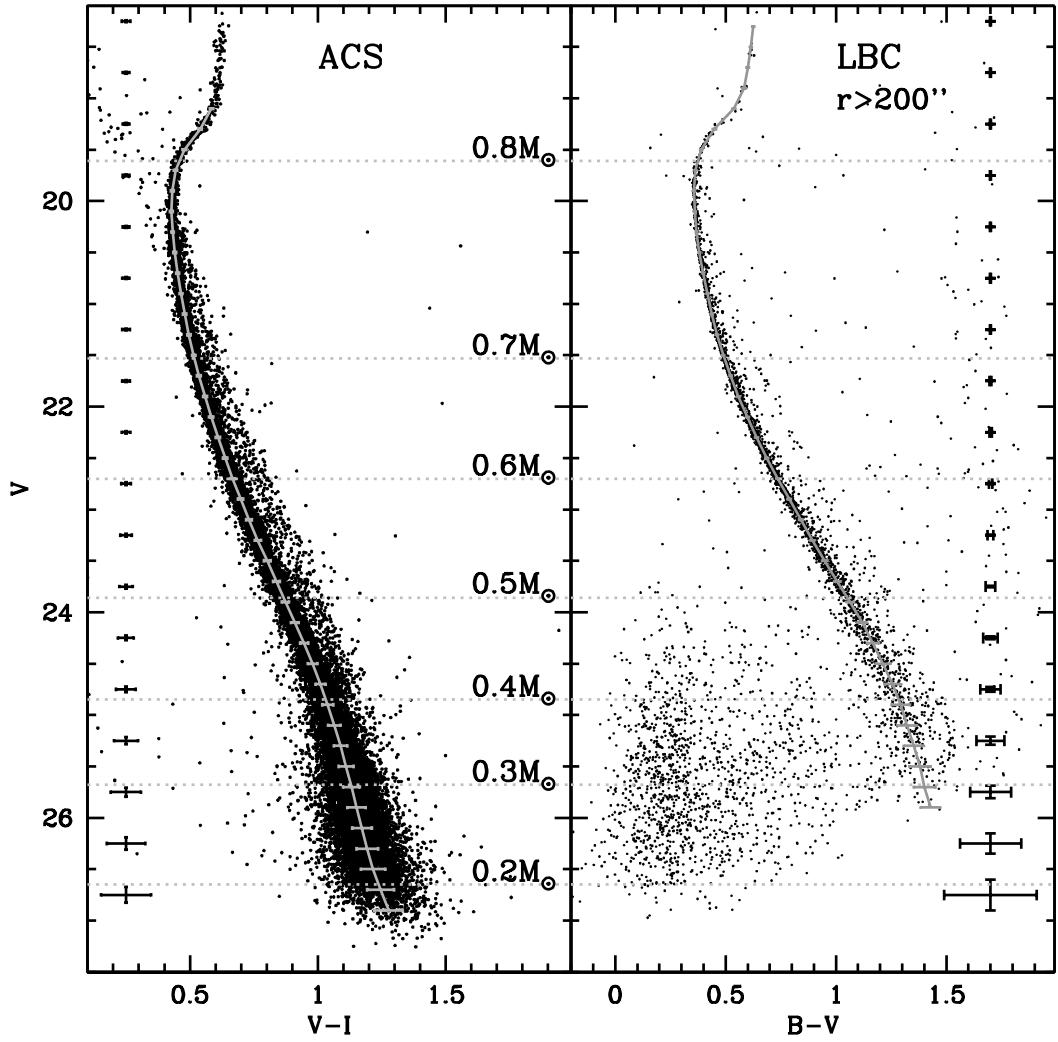} 
\caption{CMDs of the ACS and the LBC samples (left and right panels,
  respectively). The mean ridge line is shown as a solid grey line
  and it's 1$\sigma$ colour uncertainty are also marked at different
  magnitude levels.  The typical photometric errors (magnitudes and colors) for the two samples are indicated by black crosses.
  The conversion of $V$ magnitude into stellar
  masses is done using the mass-to-light law from the best-fit
  isochrone from \citet[][]{dot07}. }
\label{cmd_deep}
\end{figure}

In Figure \ref{cmd_deep} we show the color-magnitude diagrams (CMDs)
obtained with the ACS and the LBC photometric catalogs (left and right
panels, respectively).  As already discussed in Paper I, a narrow MS
is very well defined in both data-sets, witnessing the exquisite
quality of the ground-based images with respect to the ACS ones.  The
MS mean ridge lines for the two data-sets is also shown (grey solid
lines).  In the same figure we also indicate the range of stellar
masses sampled by our photometry.  The transformation of the observed
$V$ magnitudes into solar masses is done adopting the mass-to-light
law of MS stars from the the best-fit isochrone of metallicity [Fe/H]
$=-2.22$ and [$\alpha$/Fe]$=0.2$ from \citet[][]{dot07}, and assuming a distance modulus $(m -
M)_V=16.16$ and a reddening $E(B-V) = 0.0$ from \citet{f99}. The 
same model was already used in Paper I to study the radial
distribution of the BSS and binary fractions in NGC 5466.

The photometric completeness of the data-set is evaluated through the
use artificial star experiments already described in Paper I. In
short, following the recipe from \citet[][]{be02}, a number of
artificial stars with $V$ and $I$, or $V$ and $B$ magnitudes
(depending on the data-set) were randomly extracted from the observed
LFs and spatially added to the science frame in a grid of cells of
fixed width (five times larger than the typical full with at half
maximum of the stars) in order to not introduce extra stellar crowding
from the simulated stars on the images. Once the artificial stars are
added to the images, the photometric reduction is repeated
following the same strategy adopted for the original science
frames. In particular, we extracted from the catalogue of artificial 
stars the objects located inside a 2.5$\sigma$ selection box from the MS
mean ridge line. This sigma-clipping selection will be later done on the
catalogue of real star, to extract bona-fide MS  stars to be used to estimate
the cluster's LF (see Sec.~\ref{sec_lum}). 
Finally, the completeness is defined as the ratio between the
number of simulated and recovered stars in a given magnitude bin,
within the range $19 <V <27$ and a radial distance $0\arcsec <r<
1600\arcsec$ from the cluster center.

\begin{figure}
\plotone{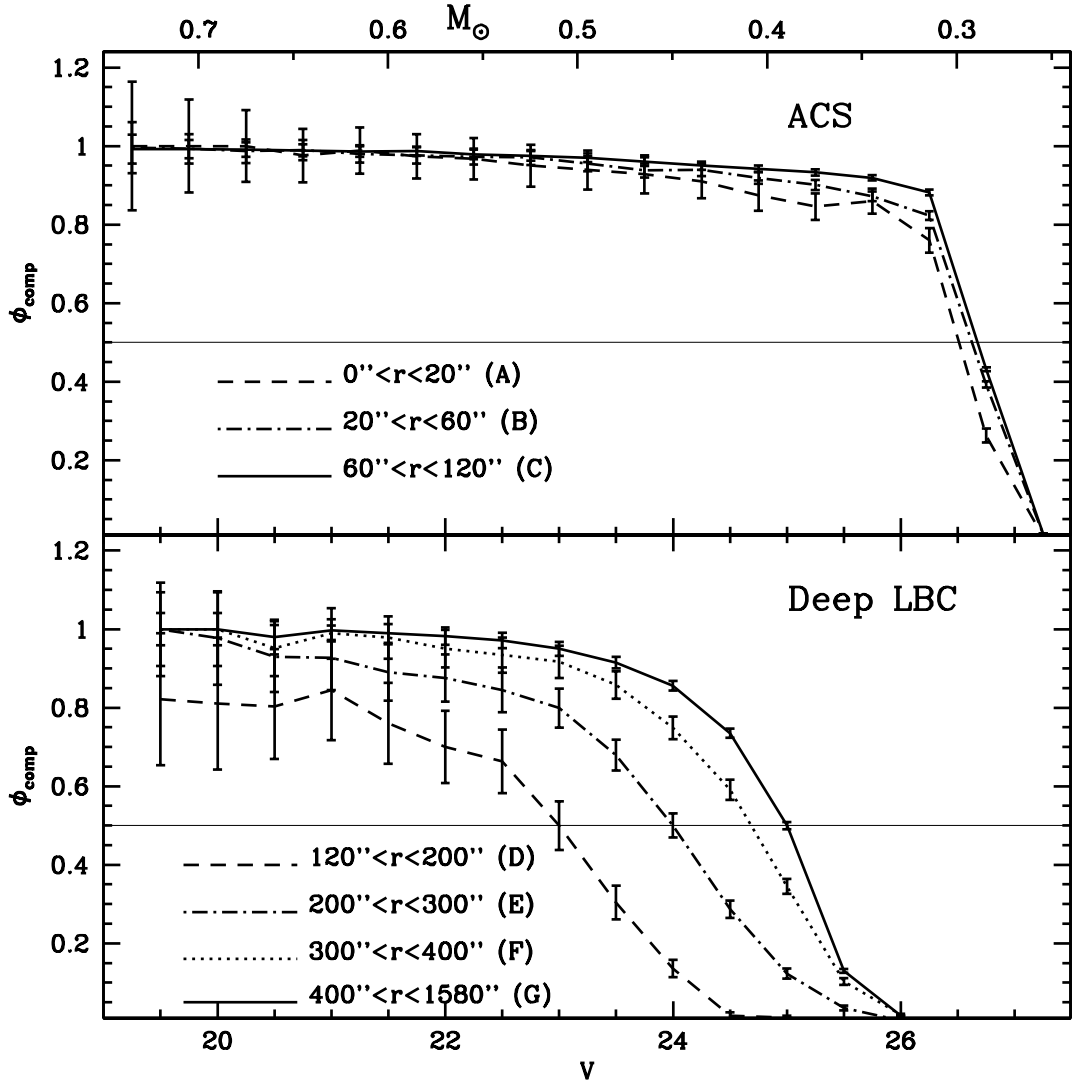} 
\caption{Photometric completeness $\phi$ as a function of the $V$
  magnitude for the ACS and the LBC data-sets divided into three and
  four concentric radial areas, respectively. The solid horizontal
  line shows the limit of 50\% of completeness. Since region D is
  characterised by a low completeness level all over the MS magnitude
  range, it has not been further considered for the LF analysis.}
\label{compl}
\end{figure}

In Figure \ref{compl} we show the photometric completeness ($\phi_{\rm
  comp}$) of the ACS and LBC data-sets (upper and lower panels,
respectively) in seven radial regions. The ACS data allow us to sample
the MS with a 50\% completeness ($\phi_{\rm comp}=0.5$) down to $V\sim
27$, corresponding to a stellar mass of $\sim0.25 M_\odot$. The
seeing-limited ground-based images are obviously more affected by
stellar crowding even in a fairly loose GC like NGC 5466. In
particular, in the first radial annulus (region D), the completeness
newer reaches the 100\% level and it drastically drops below 50\% for
stellar masses $<0.5M_\odot$. We therefore decided not to include this
radial bin in the study of the cluster's MF.  Moreover, although the
last radial bin covers a wide area, we find no significant variations
in the photometric completeness and we decided to derive a unique LF
in that region. 
We can thus assume that a single
LF/MF is well representative of the entire region. Hence, the LF and
MF of NGC 5466 are calculated in six areas, namely A, B, C, E, F and
G.

\section{Luminosity ad Mass Functions}
\label{sec_lum}
To derive the LF we first selected a catalog of bona-fide MS stars,
i.e. all stars with $V>19$ observed within 2.5$\sigma$ from the MS
mean ridge line, where $\sigma$ is the combined photometric
uncertainty in the two bands. The sample of stars selected in this
way is mostly populated by genuine single MS stars. While a fraction of
binary systems will most likely contaminate the selection, these are mostly binaries
characterised by a low mass ratios, whose mass (and light) budget 
is largely dominated by the primary star~\citep[see][]{sol07}.
The ACS and LBC catalogs obtained from
this selection contain 19,685 and 7,498 stars respectively.
The LF has been obtained by counting the number of stars with $V>19$
in steps of 0.5 mag. The lower $V$-magnitude limit in each radial
region corresponds to the value where the completeness is equal to
$50\%$.  A catalogue of simulated stars from the Galactic model of \citet[][]{rob03} 
has been used to take into account the contamination from field stars.  We properly
propagated the observer photometric uncertainties to the simulated
catalogue. The completeness- and 
field-contamination-corrected LFs determined in the six considered areas are
shown in Figure \ref{fig_mf}. The solid lines represent the
theoretical LFs obtained by multiplying a simple power-law MF of the
type $dN/dm\propto m^{\alpha}$ by the derivative of the
mass-luminosity relationship of the best-fit isochrone (see Section
\ref{sec_obs}).  With such a notation, the Salpeter IMF would have a
slope $\alpha=-2.35$, and a positive index implies that the number of
stars decreases with decreasing mass.

The best-fit models to the observed LFs have been determined by means
of a $\chi^2$ statistical test, as the ones with the MF slopes
yielding the lowest reduced-$\chi^2$. We generated a large number of
theoretical LFs using MF power-law indexes ($\alpha$) ranging from -2
to 2, in steps of 0.1.  The star counts in the magnitude range
$20<V<23.5$ have been used to normalize the theoretical LFs to the
observed one in each annulus, while the fit was performed down to
$V=25.5$ (i.e. $M>0.32 M_\odot$) in all cases.  The choice of such a
lower luminosity limit is mostly driven by the need of a compromise
between sampling the largest range of stellar masses, while dealing
with regions characterized by very different photometric
completeness. In this sense, the best-fit models in regions E and F
are partially extrapolated (in the lowest masses regime), while the
LBC observations in region G (the most external one) sample a mass
range that is perfectly compatible with the one covered by the ACS
data.

The MF power-law indexes corresponding to the best-fit LF models for
the six regions are, from the central area (A) to the most external
one (G): $\alpha=- 0.6, -0.8, -0.8, -0.9, -1.2, -1.9$, with
uncertainties of the order of 0.05 (see labels in Fig. \ref{fig_mf}).
This is the first time that the LF and the MF have been determined
over the entire radial extension of NGC 5466.  We find that the MF
slope varies only mildly from $\sim-0.6$ in the very center, to $-1.9$
out to the cluster tidal radius.
\section{Discussion}
\label{sec_disc}
\begin{figure} \centering
\includegraphics[width=0.49\textwidth]{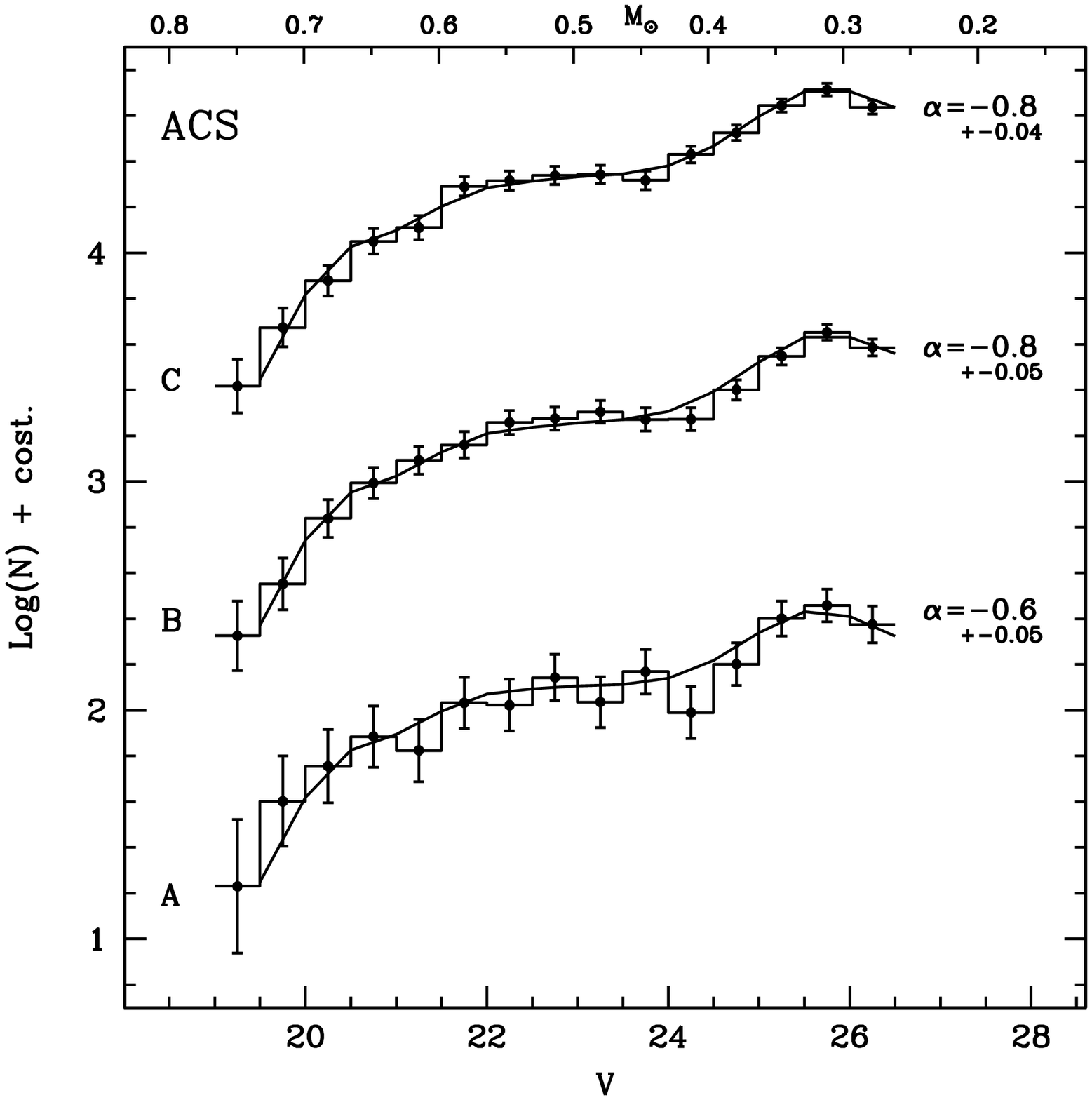}
\includegraphics[width=0.49\textwidth]{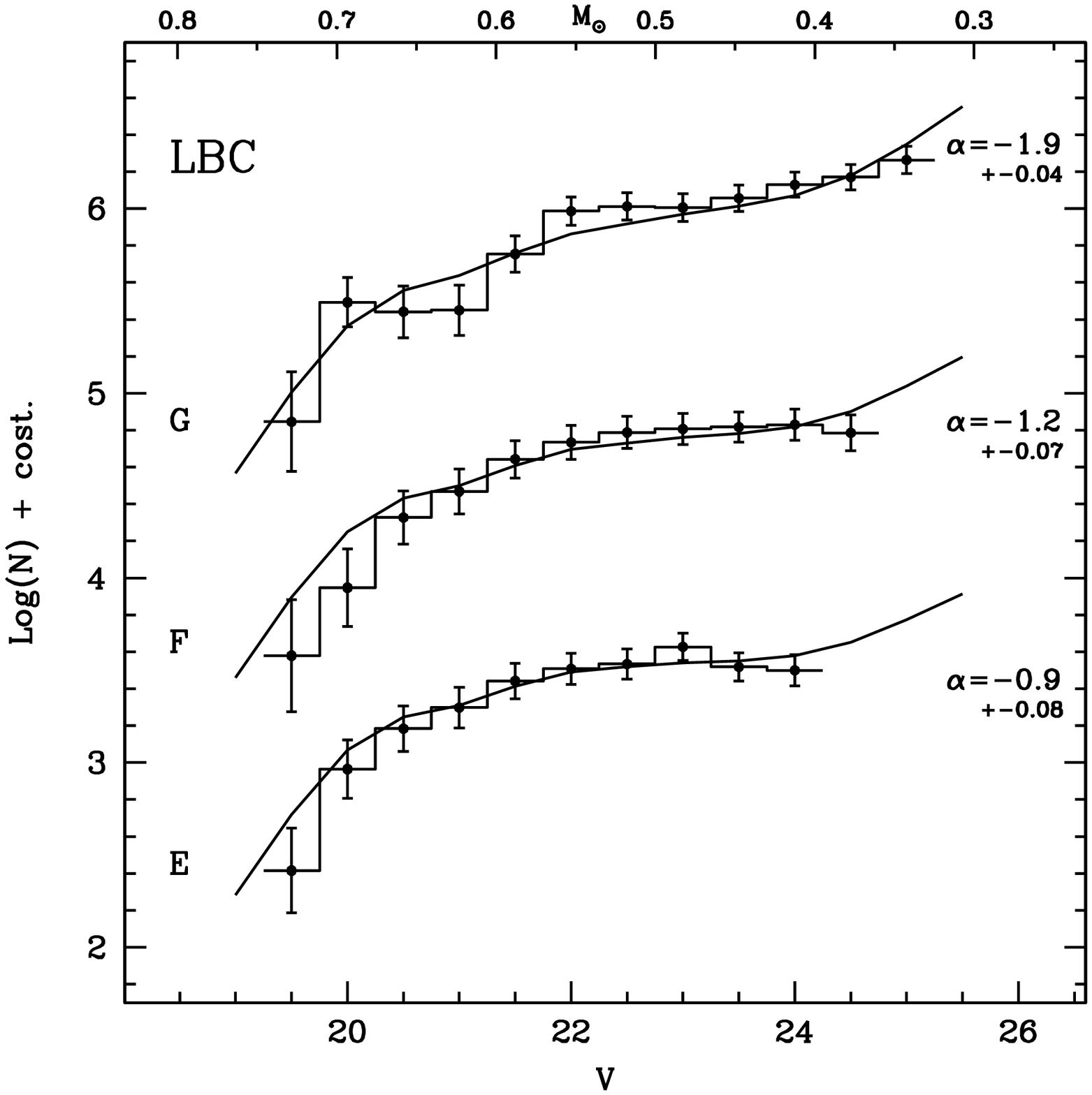}
\caption{Observed LFs of NGC 5466, as obtained in the six radial areas
  (see labels) defined in Figure \ref{compl}, from the ACS (left
  panel) and the LBC (right panel) data-sets.  The LFs are shifted by
  an arbitrary amount to make the plot more readable. The theoretical
  LFs that best-fit the data are shown as solid lines.  The
  corresponding power-law indexes of the MF are marked in the
  figure. The top axis indicates the stellar masses corresponding to
  the observed $V$ magnitudes.}

\label{fig_mf}
\end{figure}

As shown in the left-hand panel of Fig. \ref{fig_mf}, the slope of the
MF is essentially constant within the HST FOV (which corresponds to
almost twice the core radius, $r_c=72\arcsec$; \citealp{mio13}), with
only a very mild increase in the innermost region. It seems to stay
constant even out to $\sim 300\arcsec$ (although the area between
$120\arcsec$ and $200\arcsec$ could not be investigated with the
available data-sets), a distance sampled by region E, which also
includes the cluster half-mass radius \citep[$r_{\rm hm}=214\arcsec$;
  see][]{mio13}.

Quite interestingly, the MF slope measured in region E ($\alpha=-
0.9\pm 0.08$) is in very good agreement with the value of the
\emph{global} MF\footnote{Following \citet{de00}, we define the global
  MF as the present-day mass distribution of all cluster stars
  resulting from stellar evolution only, neglecting any variations due
  to the dynamical evolution of the system.}  index predicted by the
relation between $\alpha$ and the central concentration parameter $c$
shown in Figure 1 of \citet[][]{de07}.  In fact, NGC 5466 has a
concentration parameter $c=1.31$ \citep{mio13}, corresponding to a
global MF index $\alpha\sim-0.8$ (see Fig. 1 in \citealp{de07}). Such
an agreement is indeed expected, since the actual MF near the
half-mass radius should be only marginally affected by mass
segregation and should therefore be representative of the global MF of
the cluster \citep[see][and references therein]{de00,be10}.  On the
other hand, these values are not compatible with the global MF index
$\alpha=-1.15\pm 0.03$ estimated by \citet{pau10}, who used only ACS data 
and corrected the MF for the effects of mass segregation using a set of
multi-mass King models.

\begin{figure}
\plotone{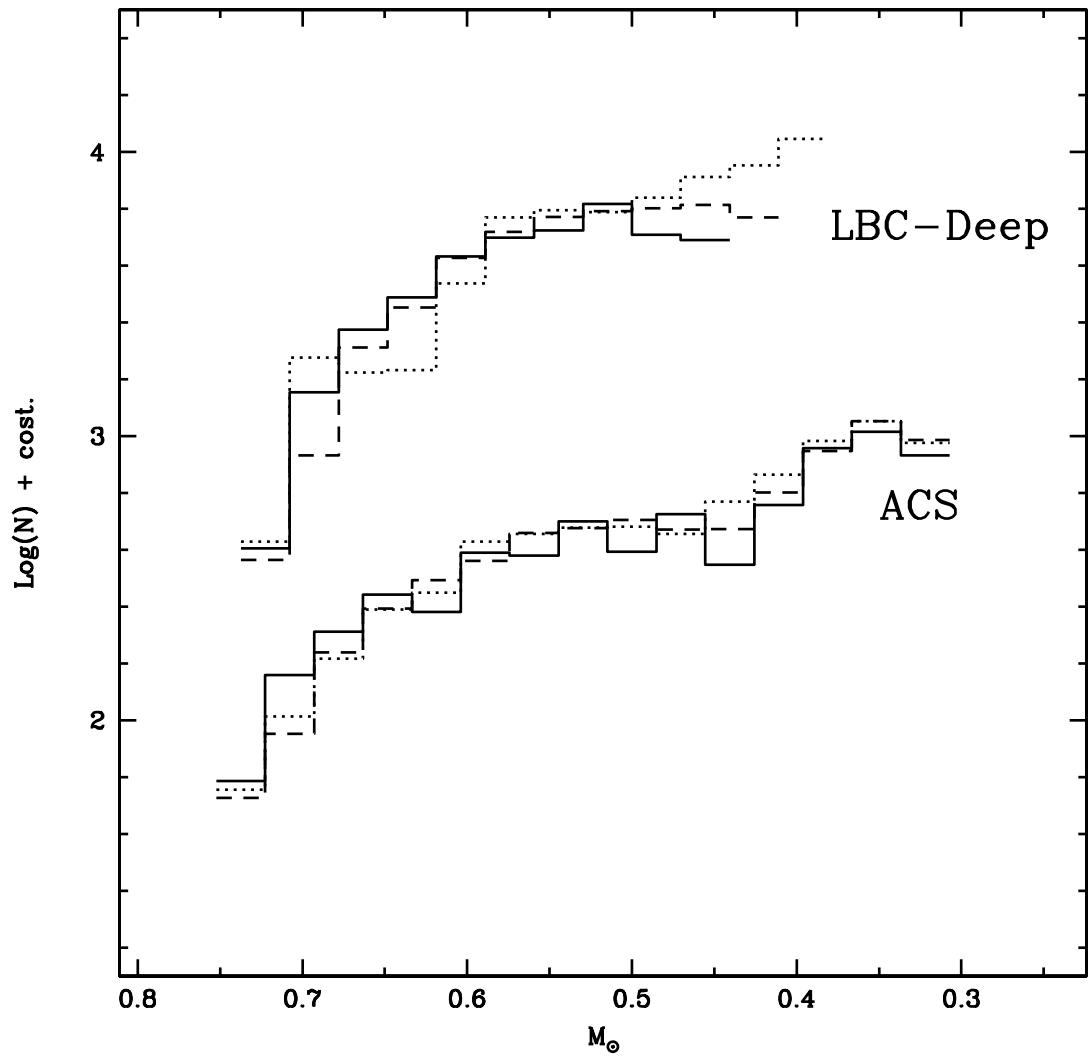} 
\caption{Direct comparison among the observed LFs in the
  ACS and in the LBC data-sets. For both data-set, a solid line
  represents the most internal region, a dashed line shows the region
  at intermediate distance while the dotted line is used for the one located
  at larger distance respect to the others in the same data-set.}
\label{fig_LF}
\end{figure}

As we move outwards, $\alpha$ (slowly) decreases. This can be also
appreciated in Figure \ref{fig_LF}, where the LF measured in the
different regions of the ACS and the LBC FOVs are over-plotted. In
the areas sampled by the ACS data (A, B and C), the shape of the LF is
the same, while in the regions covered by the LBC observations (E, F
and G) the most external LF (dotted line) shows an excess in the
star counts at the low-mass end, with respect to the LFs measured in
the other two regions (solid and dashed lines). Such a behaviour in
the most external bin is in agreement with what expected for a cluster
moderately affected by mass segregation.

The conclusion in favour of a cluster not heavily affected by dynamical
evolution (i.e., a dynamically-young system) is also supported by the
comparison with the results that we obtained from a similar analysis
performed in the GC M10. In \citet{be10} we studied the radial
distribution of the MF slope of M10 in the same mass range considered
here, and out to 5 times the cluster core radius \citep[$r_c=41\arcsec$,
or $2r_{\rm hm}$][]{mio13}. We found that, from the center out to $5 r_c$,
the MF index varies as: $\alpha=0.7, 0.4, 0.1, -0.3, -0.6, -0.9$.
Hence, the gradient of $\alpha$ in M10, defined as the absolute value
of difference between its central value and the one at $5 r_c$, is
$\Delta\alpha_{5r_c}=1.6$.  In the case of NGC 5466, the same quantity
can be calculated as the difference between the value of $\alpha$ in
region A ($\alpha=-0.6$,) and that measured in region F
($\alpha=-1.2$), which includes the $r=5\times r_c$ (approximately $2 r_{hm}$)
distance from the cluster center.  Hence, in the case of NGC 5466 we obtain
$\Delta\alpha_{5r_c} = 0.6$, meaning that the variation of the MF
slope is more than a factor of 2 shallower in NGC 5466 than in M10.
Such a comparison indicates that two-body relaxation processes worked
more efficiently in shaping the mass distribution of MS stars in M10,
with respect to what happened in NGC 5466. 

\begin{figure}
\plotone{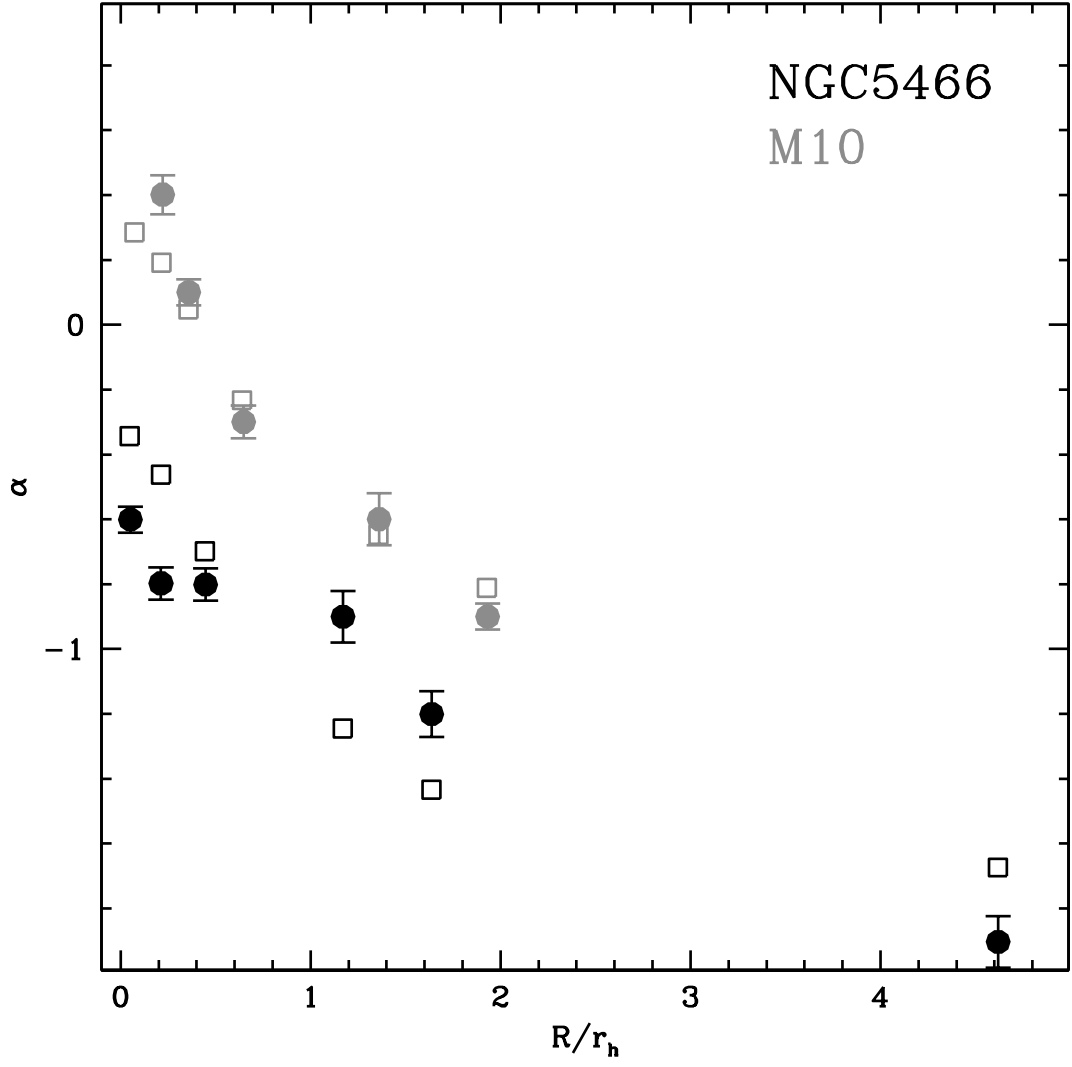} 
\caption{Radial distribution of the 
observed slope $\alpha$ of the MFs for NGC~5466 and M10 (black and grey solid circles)
with respect to the theoretical ones obtained in the same radial regions (open squares) 
by using a set of multi-mass King-Michie models. The uncertainties on the observed values
of $\alpha$ are also shown.
}
\label{comp}
\end{figure}

Note that the two clusters have quite different concentrations and global MF~\citep[][]{mio13, be10}. 
So, it is possible that this difference is reflected in the observed  $\Delta \alpha$ variation. To 
test this possibility, for both clusters we fit the surface brightness and the MF in the radial bin containing the 
half-mass radius (where the MF slope should resemble the actual slope of the global MF)
using a set of multi-mass King-Michie models~\citep[][]{gg79}. These models have been constructed 
assuming 8 mass bins ranging from 0.1 $M_{\odot}$ to the mass at the Red Giant Branch (RGB) tip 
and standard assumptions  on the fraction of remnants~\citep[see][]{so12}. The theoretical MF slopes have been 
then estimated in the same radial bins adopted in the observations and are shown in Fig.~\ref{comp} (open
squares) together with the observed ones (solid circles). 
While the observed variation of $\alpha$ in M10 (grey solid circles) is slightly steeper than what predicted by models (grey open squares), 
the opposite is happens for NGC~5466 which appears to be under-segregated with respect 
the best-fit model in the same radial range. 
It is important to remark here that, according to~\citet[][]{mil12},
the core binary fraction in NGC 5466 is $\sim14\%$, while it is only $\sim8\%$
in M10. In~\citet[][]{be10}, through a set of realistic
N-BODY simulations, we showed that binaries play a non negligible 
role in the dynamical evolution of the cluster, acting as energy source
that quenches the effect of mass segregation in a cluster. Hence, the difference
in the binary fraction in the two clusters can be at the origin (at least partially) of the
different behaviour between the observed and the predicted radial distribution profiles of $\alpha$.
Nonetheless, the models shown in Fig.~\ref{comp} account for the different concentrations 
and MFs, since the same recipe for mass segregation is adopted. Hence the opposite systematic deviations that the observed data show with respect to the 
predicted behaviours indicate an actual difference in the efficiency of mass segregation which appears 
to be stronger in M10 than in NGC~5466.

This result is per se not a surprise. In fact, adopting the cluster's structural
parameter from~\citet[][]{mio13} and the equations 10 and 11 from
\citet[][]{dj93}, we calculate the clusters' core and half-mass 
relaxation times ($t_{rc}$ and $t_{rh}$) to be 2.6 and 19.9 Gyr for NGC 5466,
and 0.3 and 1.9 Gyr for M10, respectively. Hence, the comparison of the
expected evolutionary time-scales already suggests that NGC~5466 should be
dynamically younger that M10. The study proposed in this paper
offers a new observational proof of the different dynamical states of the two clusters.

Very interestingly, the same conclusion is obtained also from the
study of the radial distribution of BSSs and binary systems 
(see Paper I for NGC 5466, and \citealp{da11,dale13} for
M10). In particular, \citet{fe12} showed that the radial distribution
of BSSs, normalized to that of a ``normal'' stellar population taken as
proxy of the distribution of the cluster's stars, is an efficient
tracer of the dynamical evolution of the hosting cluster. In fact, the
position of the minimum ($r_{\rm min}$) of the BSS radial distribution
marks the distance at which dynamical friction has already been
effective in segregating BSSs towards the cluster centre.  Through the
comparison among the values of $r_{\rm min}$ measured in several
Galactic GCs, \citet{fe12} defined the so-called ``dynamical clock'',
an empirical tool able to rank GCs according to their dynamical age.
In this context, in Paper I we have shown that $r_{\rm min}$ is
$\sim2.5r_c$ in NGC 5466, while it corresponds to $\sim10 r_c$ in M10
\citep{dale13}. According to the dynamical clock, NGC 5466 is
therefore classified as an ``early Family II'' GC, while M10 is ranked
in the ``evolved Family II'' sub-class, where dynamical friction has
been more effective in shaping the radial distribution of BSSs.

All the three different and independent dynamical indicators studied
so far (namely, the radial trend of the MF slope presented in this
work, and the radial distributions of BSSs and binaries studied in
Paper I) therefore agree upon showing that NGC 5466 is a GC that just
started to evolve dynamically.

\acknowledgments{We wish to thank the anonymous referee for insightful 
 comments that have helped to improve the presentation of our work.
 This research is part of the project COSMIC-LAB
  funded by the European Research Council (under contract
  ERC-2010-AdG-267675). This research used the facilities of the 
  Italian Center for Astronomical Archive (IA2) operated by INAF at 
  the Astronomical Observatory of Trieste. Also
  based on observations made with the NASA/ESA Hubble Space Telescope,
  obtained from the data archive at the Space Telescope
  Institute. STScI is operated by the association of Universities for
  Research in Astronomy, Inc. under the NASA contract NAS 5-26555.  }

\clearpage




\clearpage








\clearpage
\begin{table}
\label{tab_obs}
 \centering
  \caption{Log of the observations}
  \begin{tabular}{@{}ccccc@{}}
  \hline
   Data-set     & Number of exposures & Filter & Exposure time & Date of observations\\
            &                 &        &     (s) & \\
 \hline
\multicolumn{5}{c}{Deep Sample}\\
 \hline
 LBC-blue  & 11 & $B$ & 400 & 2010-04-11\\
      & 15 & $V$ & 200 & 2010-04-11\\
  ACS   & 5 & $V_{606}$ & 340 & 2006-04-12\\
       & 5 & $I_{814}$ & 350 & 2006-04-12\\
\hline
\end{tabular}
\end{table}


\end{document}